\documentclass[12pt]{article}
\usepackage{hyperref}
\usepackage{graphicx}
\usepackage{epsfig}
\usepackage{times}
\usepackage{amsmath,amssymb}
\newcommand{\be}{\begin{equation}}
\newcommand{\ee}{\end{equation}}
\newcommand{\bea}{\begin{eqnarray}}
\newcommand{\eea}{\end{eqnarray}}
\newcommand{\bref}[1]{(\ref{#1})}

\begin{document}
\begin{flushright}
RITS-PP-002
\end{flushright}
\vskip 1cm
\begin{center}
{\Large\bf de Vaucouleurs-Ikeuchi Diagram and Commutation Relations 
among Phase Space Coordinates}
\vskip 1cm
{Takeshi FUKUYAMA \footnote{E-mail: fukuyama@se.ritsumei.ac.jp}
and Tatsuru KIKUCHI \footnote{E-mail: rp009979@se.ritsumei.ac.jp}
}\\
\medskip
{\it Department of Physics, Ritsumeikan University,\\
Kusatsu, Shiga, 525-8577, Japan}
\end{center}
\begin{abstract}
We consider the relations between de Vaucouleurs-Ikeuchi diagram and 
generalized commutation relations among the coordinates and momenta.
All physical objects in the Universe ranging from elementary particles
to super cluster of galaxies are confined within the Triangle of 
the de Vaucouleurs-Ikeuchi diagram on the matter density versus scale 
length plane. These three boundaries are characterized by
the quantum uncertainty principle, gravitational event horizon, 
and cosmological constant. These are specified by the non-zero 
commutation relations $[x_\mu,p_\nu],~[x_\mu, x_\nu]$ (strictly $[x_i,t]$) 
and $[p_\mu,p_\nu]$, respectively.
The canonical commutation relation $[x_i,p_j]$ are slightly modified, 
which preserves the self consistency as a whole.
\vskip 1cm
\end{abstract}
\section{Introduction}
All physical entities ranging from microscopic quark-leptons to 
macroscopic super cluster of galaxies are confined within the Triangle 
in the de Vaucouleurs-Ikeuchi diagram \cite{Ikeuchi} on log-log scale 
of matter density versus cosmological length (Fig.1).

%%%%%%%%%%%%%%%%%%%%%  de Vaucouleurs-Ikeuchi diagram %%%%%%%%%%%%%%%%%%%%%%
\begin{figure}[h]
\begin{center}
\epsfig{file=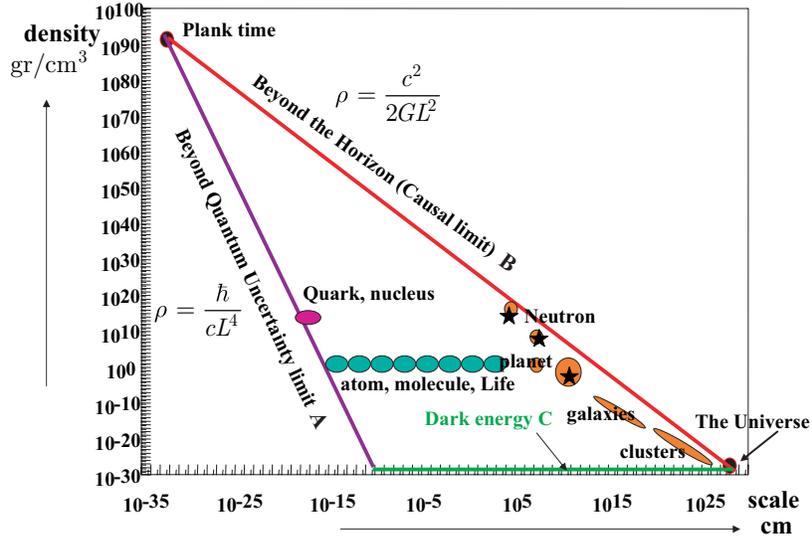, width=12cm}
\end{center}
\caption{The de Vaucouleurs-Ikeuchi diagram and the Triangle}
\end{figure}
%%%%%%%%%%%%%%%%%%%%%%%%%%%%%%%%%%%%%%%%%%%%%%%%%%%%%%%%%%%%%%%%%%%%%%%%%%%%%%
The line A corresponds to the `quantum limiting line' beyond 
(toward lower left) which all the structures are perturbed by 
quantum fluctuations and cannot exist as  observable bodies. 
The limiting line is given as follows: 
the Compton length associated with a particle of mass $M$
\bea
\lambda  = \frac{h}{{Mc}}
\eea
which is the smallest de Broglie wave length associated with matter 
$\lambda  = \frac{h}{{Mv}}$. 
The corresponding mass density is given by
\bea
\rho_{{\rm quantum}} 
= \frac{M}{{\left({4\pi/3} \right) \lambda^3 }}  
= \frac{h}{{\left({4\pi /3} \right) c L^4 }} \;.
\eea
The line B corresponds to the `causal limiting line' beyond 
(toward upper right) which all the structures collapse gravitationally 
and form black holes which have no individuality. The limiting line 
is given by considering gravitational (Schwarzshild) radius
\bea
L = \frac{{2GM}}{{c^2 }} \;.
\eea
The limiting mass density is obtained by 
\bea
\rho_{{\rm causal}} 
= \frac{M}{{\left({4\pi /3} \right) L^3 }}
= \frac{{c^2 }}{{\left({4\pi/3} \right) G L^2 }} \;,
\eea
and therefore inversely proportional to the length squared. 

The horizontal line C corresponds to the cosmological constant 
($\rho_{\rm const}$) \cite{wmap} below which no mass density is attainable.

All observable constituents in the Universe are within the triangle 
enclosed by these three edges. The first uncertainty principles are 
the realization of
\be
[x_\mu,p_\nu]=i\hbar \eta _{\mu\nu}
\label{xpc}
\ee
and leads us to the quantum uncertainty principle
\be
\Delta x\Delta p\geq \hbar,~~\Delta t\Delta E\geq \hbar \;.
\label{xp}
\ee
(Here and hereafter we use natural units except for the case to emphasize 
$c$ or $\hbar$.) As we will show, these relations are slightly modified 
in the total framework. Before discussing the line B we proceed to study 
the line C ($\rho_{\rm const}$), which is the dark energy or 
cosmological constant.
\be
\rho_{\rm const}=10^{-122}M_{\rm Pl}^4 \;.
\label{cc}
\ee
The cosmological constant is formulated in de Sitter invariant gravity
\cite{mansouri}. The Einstein-de Sitter gravity Lagrangian
\be
{\cal L}=\partial _\mu K^\mu -\frac{1}{16\pi G_N}
\sqrt{-g} \left(R + \frac{6}{\ell^2} \right)
\label{einstein}
\ee
which is invariant under the de Sitter gauge group
\be
\eta_{AB} X^A X^B = \ell^2 ~(A,B=0,.,4) \;.
\label{desitter}
\ee
Hence the cosmological constant is given by
\be
\rho_{\rm const} = \frac{3}{8\pi \ell^2G_N} \;.
\ee
The inflationary coordinates $(t, x^i)$ are defined by
\bea
&&
X^0 = \ell \sinh (t/\ell) + \frac{(x^i)^2}{2 \ell} e^{t/\ell} \;,
\nonumber\\
&&
X^i = x^i e^{t/\ell} \;,
\nonumber\\
&&
X^4 = \ell \cosh (t/\ell) - \frac{(x^i)^2}{2 \ell} e^{t/\ell}\;.
\nonumber\\
\eea
This leads to the metric in the inflationary era,
\be
ds^2 = - dt^2 + e^{2 t /\ell} (dx^i)^2\;.
\ee
Hence the size of the Universe grows as $\sim e^{2 t /\ell}$. 
On the other hand, the static metric is given by
\be
ds^2 = - \left(1 - \frac{r^2}{\ell^2}  \right) dt^2
+ \frac{dr^2}{\left(1 - \frac{r^2}{\ell^2}  \right)}
+ r^2 d \Omega_2^2 \;.
\ee
Note that the line element becomes singular for $r = \ell$, 
which is a cosmological event horizon to which we can associate 
an entropy
\be
S = \frac{ 2 \pi \ell^{2}}{G_N} \;.
\ee
The Penrose diagram is shown in Fig.~2.
%%%%%%%%%%%%%%%%%%%%%  Penrose diagram %%%%%%%%%%%%%%%%%%%%%%%%%%%%%%%%
\begin{figure}[h]
\begin{center}
\epsfig{file=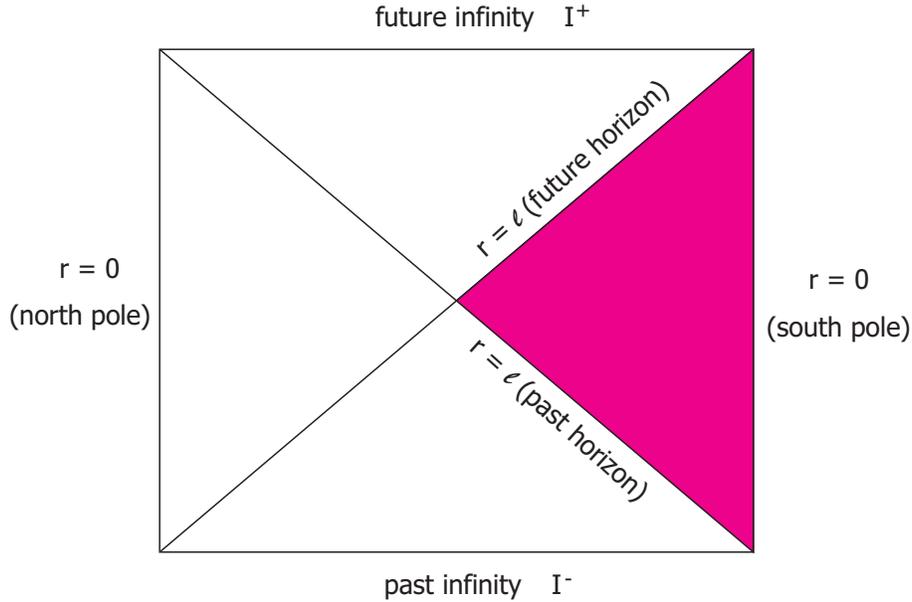, width=12cm}
\end{center}
\caption{The Penrose diagram for de Sitter space \cite{hawking-ellis}.}
\end{figure}
%%%%%%%%%%%%%%%%%%%%%%%%%%%%%%%%%%%%%%%%%%%%%%%%%%%%%%%%%%%%%%%%%%%%%%%
The peculiar feature of de Sitter space is that no single observer can
acccess the entire space-time. In the diagram, an observer sitting on 
the south pole will never be able to observe anything past the diagonal 
line stretching from the north pole at ${\cal I}^-$ to the south pole 
at ${\cal I}^+$, and the shaded region is called as ``causal diamond''.
For more details, see, e.g. \cite{strominger}.

The generators of de Sitter group transformation satisfy 
the commutation relation
\be
[M_{AB},M_{CD}] = i \hbar 
(\eta_{AC}M_{BD}+\eta_{BD}M_{AC}-\eta_{AD}M_{BC}-\eta_{BC}M_{AB})
~~(A,..,D=0,..,4) \;.
\label{desitter}
\ee
Momenta $p_\mu ~(\mu=0,..,3)$ are defined by
\be
p_\mu =M_{\mu 4}/\ell
\ee
and they are not commutative to each others,
\be
[p_\mu,p_\nu] = i \hbar \ell^{-2} M_{\mu\nu} \;.
\label{ppc}
\ee
It should be remarked that \bref{desitter} reduces to the Poincar\'e 
algebra in the Wigner-In\"on\"u contraction $\ell\rightarrow \infty$.
\par
Thus the line A and the line C are characterised by 
$[x_\mu,p_\nu]\neq 0$ and $[p_\mu,p_\nu]\neq 0$, respectively.
So we are attracted to think that the remaining line B reflects 
new physics related with $[x_\mu,x_\nu]\neq 0$.
We will show in the following that this is indeed the case.

At first we may postulate
\be
[x_i,x_j]=i\theta_{ij}~(i,j=1,2,3)
\label{ncg}
\ee
as in the non-commutative geometry \cite{nekrasov}.
However, we can not associate this with event horizon, though its 
possibility can not be refused. Instead, we adopt the Yoneya's relation 
in string theory \cite{yoneya}.
\be
[x_i,t]=\ell_s^2 = \alpha' \;,
\label{xtc}
\ee
where $\ell_s$ is the string length.  This is necessary condition 
to be free from ultra-violet divergence and can be interpreted also 
by considering the scattering of two D-particles with impact parameter 
of order $\Delta X\geq \ell_s$ and obtain
\be
\Delta X\Delta T\geq \ell_s^2 \;.
\label{up}
\ee
As was shown in \cite{seiberg} \bref{xtc} in string theory is free from 
the acausal mode unlike the same case of the simple non-commutative 
geometry \bref{ncg}. The implications of the stringy space-time uncertainty 
\bref{up} on the spectrum of cosmological fluctuations was studied in 
\cite{ho}.

Considering the event horizon of the size of $\ell_s$, we can obtain 
the bonus to reproduce the Bekenstein-Hawking entropy $S_{\rm BH}$ relation 
\cite{yoneya2} from microscopic sum of quantum states,
\be
\mbox{log}~d(E)
\approx G_N^{\frac{1}{D-3}}E^{\frac{D-2}{D-3}}
\approx S_{\rm BH}
\equiv \frac{\mbox{Area of BH}}{4G_N} \;.
\ee
Here $d(E)$ is a quantum degeneracy, 
\be
d(E)=\mbox{exp}\left(2\pi\sqrt{(D-2)/6}~ \ell_s E \right) \;,
\ee
and $D$ is the space-time dimension.

Now using the recent advances in string theory of AdS/CFT correspondence
we can know about a connection among two scales $\ell$ and $\ell_s$. 
That is, there is a duality between the ${\rm AdS}_5 \times {\rm S}_5$ 
compactification of the type IIB string theory and the superconformal 
Yang-Mills theory of the ${\rm SU}(N)$ gauge group \cite{AdS/CFT}. Thus 
the de Sitter radius $\ell$ is given in terms of the string scale 
$\ell_s$ and the gauge index $N$ as
\be
\ell = (4 \pi g_s N)^{1/4} \ell_s \;,
\ee
where $g_s = \sqrt{8 \pi G_N/\alpha'}$ is the string coupling constant. 
For the more detailed argument of the de Sitter entropy,
see, e.g. \cite{horava}.

So far we have considered the extended canonical commutation relations. 
Here we argue on the consistency of those relations. We assume that 
the Jacobi identity still holds in this extended commutation relations. 
We consider
\be
[[p_i,p_j],x_k]+[[x_k,p_i],p_j]+[[p_j,x_k],p_i]=0 \;.
\ee
It goes from \bref{ppc} that the first term is
\be
[[p_i,p_j],x_k]=i\hbar \ell^{-2}[M_{ij},x_k]
=-\hbar \ell^{-2}(\delta _{ik}x_j-\delta _{jk}x_i) \;.
\label{jacobi1}
\ee
Hence to preserve the Jacobi's identity we must extend the canonical 
commutation relation as
\be
[x_i,p_j]=i\hbar \delta_{ij}(1+\beta x^2) \;.
\label{ccr}
\ee
Here $\beta$ is a tiny parameter to be constrained from 
\bref{jacobi1} as
\be
\beta=-\frac{1}{2 \ell^2}
\label{lbeta}
\ee
The different extension of $[x_i,p_j]$ from ours was considered in 
\cite{kempf},
\be
[x_i,p_j] = i \hbar \delta_{ij}(1+\beta'p^2)
\ee
which leads to the minimal length 
$\Delta x_{\rm min} \geq \hbar \sqrt{\beta'} \approx \ell_s$. 
So different tiny corrections to canonical commutation relations 
induces enormously different length scale $\ell$ and $\ell_s$.
If we consider another type of the Jacobi identity
\be
[[x_i,p_j],x_k]+[[x_k,x_i],p_j]+[[p_j,x_k],x_i]=0 \;.
\label{jacobi2}
\ee
\bref{jacobi2} is reduced from \bref{ccr} to
\be
\beta\delta_{ij}[x^2,x_k]+\beta\delta_{jk}[x^2,x_i]+[[x_k,x_i],p_j] = 0 \;.
\ee
This does not allow the extension of
\be
[x_i,x_j]=0 \;.
\ee
It should be remarked that the above arguments do not restrict 
$[t,x]$ commutation relation. These commutation relations allows 
the representation
\be
\hat{x}_i = x_i,~~~
\hat{p}_j = -i\hbar (1+\beta x^2)\frac{\partial}{\partial x_i}
\ee
unlike \bref{ncg}.

The commutation relation $[p_i,p_j]$ leads to \bref{lbeta} again.
If we consider \cite{landau}
\be
\int \left|\alpha x\psi~+(1+\beta x^2)\psi \right|^2 dx \geq 0
\ee
for arbitrary real $\alpha$, we obtain the generalized uncertainty relation
\be
\Delta p\geq \frac{\hbar}{2}(\frac{1}{\Delta x}+3\beta \Delta x),
\ee
which implies the minimum momentum
\be
\Delta p_{\rm min}\geq \frac{\hbar}{\ell} \;.
\ee
Using the generalized Poisson bracket $\{A,B\}_{\rm P.B.}$ 
obtained from the correspondence,
\be
\frac{[A,B]}{i\hbar}=\{A,B\}_{\rm P.B.}
\ee
we can show that the Liouville theorem holds,
\be
\frac{d^{D-1}x' d^{D-1}p'}{(1+\beta x'^2)^{D-1}}
=\frac{d^{D-1}x d^{D-1}p}{(1+\beta x^2)^{D-1}}
\ee
thanks to \bref{lbeta}. Obviously this does not serve as 
the suppression of cosmological constant like \cite{okamura}. 
However the correct magnitude of cosmological constant is guaranteed 
as the lowest energy of de Sitter invariant vacuum \cite{mansouri}. 

Unfortunately we can not find the bridge between the cosmological constant 
and the string length scale. One of the reasons is that the ordinary 
commutation relations are concerned the spatial coordinates of phase space. 
On the other hand, the string length scale appeared in the commutation 
relation between space and time coordinates. 

Consequently we have discussed the Triangle in the de Vaucouleurs-Ikeuchi 
diagrams. The three boundaries characterise quantum mechanics, black hole, 
and cosmological constant, which are described by the non-zero 
commutation relations, $[x_i,p_j],~[t,x_i],$ and $[p_i,p_j]$, respectively. 
The first canonical commutation relation is modified to 
$[x_i,p_j]=i\hbar \delta_{ij} (1+\beta x^2)$, which preserves 
the self consistency as a whole. This is the peculiar property 
of de Vaucouleurs-Ikeuchi diagram (log density-log length diagram).

\medskip
\vskip 5 mm
Acknowledgments\\
T.F. is grateful to M.~Morikawa for useful comments and helps. 
He also thanks C.Q.~Geng at NCTS for discussions and High Energy group 
at NCTS and Theory Division at KEK for hospitality. 
This work is supported by the Grant-in-Aid for Scientific Research 
from the Ministry of Education, Science and Culture of Japan (\#16540269). 
The work of T.K. is supported by the Research Fellowship of the Japan 
Society for the Promotion of Science (\#7336).
\vspace{5 mm}
\par

\end{document}